\documentclass[twocolumn,pra,showpacs,preprintnumbers]{revtex4}

\usepackage{graphicx}
\usepackage{amsfonts,amssymb, amsmath}
\usepackage{wasysym}  
\usepackage{hyperref}

\graphicspath{{epsfigs/}{pdffigs/}}

\newcommand{\mvec}[1]{{\boldsymbol{ #1}}}  
\newcommand{\subm}{ \prec^w }
\newcommand{\pisq}{{\pi_{\square}}}
\newcommand{\pisqb}{{\pi_{\square\text{b}}}}
\newcommand{\pibt}{\pi_{\bowtie}} 
\newcommand{\thetakag}{\theta_{\text{kag}}}
\newcommand{\iC}{{\cal C}}

\newcommand{\ket}[1]{\,|#1 \rangle}
\newcommand{\crt}{{\cal S}_0}
\newcommand{\cfr}{{\cal S}}

\newcommand{\ourrefs} {Refs.~\onlinecite{Natphys.3.1745} and \onlinecite{perseguers:022308}}

\newcommand{\widthone}{0.85\columnwidth}
\newcommand{\widthtwo}{\columnwidth}

\newcommand{\figr}[1]{Fig.~\ref{#1}}

\newcommand{\tabr}[1]{Table~\ref{#1}}

\newcommand{\secr}[1]{Sec.~\ref{#1}}

\newcommand{\ssecr}[1]{Sec.~\ref{#1}}

\newcommand{\appendr}[1]{Appendix~\ref{#1}}

\newcommand{\Nexamp}{ five }

\begin{document}

\title{Enhancement of Entanglement Percolation in Quantum Networks via Lattice Transformations}

\author{G John Lapeyre, Jr.}
\email{lapeyre@physics.arizona.edu}

\noaffiliation

\author{Jan Wehr}
\affiliation{Department of Mathematics, The University of Arizona, Tucson, AZ 85721-0089, USA}

\author{Maciej Lewenstein}
\affiliation{ICFO-Institut de Ci\`encies Fot\`oniques, Mediterranean Technology Park, E-08860 Castelldefels (Barcelona), Spain\\
  ICREA-Instituci\'o Catalana de Recerca i Estudis Avan\c
  cats, Lluis Companys 23, 08010 Barcelona, Spain}

\date{March 4, 2009}

\begin{abstract}
  We study strategies for establishing long-distance
  entanglement in quantum networks. Specifically, we consider networks
  consisting of regular lattices of nodes, in which the
  nearest neighbors share a pure, but non-maximally entangled
  pair of qubits. We look for strategies that use local
  operations and classical communication. We compare the
  classical entanglement percolation protocol, in which every
  network connection is converted with a certain probability to
  a singlet, with protocols in which classical
  entanglement percolation is preceded by measurements
  designed to transform the lattice structure in a way that
  enhances entanglement percolation.  We
  analyze five examples of such comparisons between protocols and point out
  certain rules and regularities in their performance as a function of
  degree of entanglement and choice of operations.
\end{abstract}


\maketitle

\section{Introduction \label{sec:intro}}

Entanglement is the property of states of multipartite
quantum systems that is the most important resource for
quantum information processing \cite{HHH08}. For this reason,
one of the most important tasks of quantum information
science is to establish entanglement at long distances in
quantum networks, and to optimize final entanglement and
probability of success.

Quantum networks \cite{CZKM97,BBMNK07} play a key role in
quantum information processing. Here we limit our attention
to those networks in which quantum states can be prepared
initially and shared. That is, entanglement can be generated
between neighboring or, at least, not-too-widely-separated
nodes (or stations). There are two instances in which the
above-mentioned tasks become obviously relevant. On one
hand, one can consider macroscopic quantum communication
networks, such as cryptographic networks, or more generally
quantum communication nets \cite{E91,BBC+93,seqoqc}, or
distributed quantum computation \cite{CEHM99} involving
arbitrary nodes of the network. The second instance concerns
microscopic or mesoscopic networks that could constitute
architectures of quantum computers (cf.
Ref.~\onlinecite{Nielsen}).

\begin{figure}
  \includegraphics[width=\widthone]{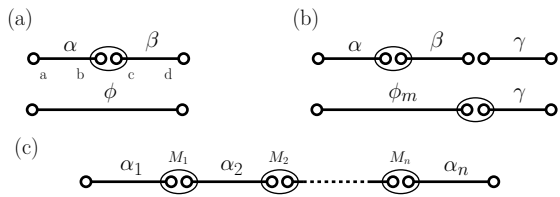}
  \caption{\label{fig:swapping} Entanglement swapping
    and 1D repeater. Circles are qubits. Heavy lines represent
    pure non-maximally-entangled states. Loops represent the
    entanglement swapping measurement on a pair of qubits.  a)
    Two states $\alpha$ and $\beta$ of the form specified in
    (\ref{stateAB}). After the operation, qubits $a$ and $d$
    may be in an entangled, mixed state. b) swapping as the
    first step in a repeater. c) a 1D chain of repeaters.  }
\end{figure}

Despite enormous progress in experimental techniques (cf.
Ref.~\onlinecite{seqoqc} and references therein), it is in
principle a very hard task to establish entanglement at
large distances due to decoherence and attenuation effects.
Two remedies for this problem have been proposed:
\begin{itemize}
\item {\it Quantum repeaters}. This concept has been
  developed for 1D quantum communication chains
  \cite{BDCZ98,DBCZ99,CTSL05,HKBD06}. Although the simple
  entanglement swapping \cite{ZZHE93} procedure can lead to
  quantum communication at large distances (see
  \figr{fig:swapping}), for imperfect resources, the
  performance of such communication chains decays
  exponentially with the distance (i.e. the number of
  repeaters).  However, one can use more sophisticated
  quantum repeater protocols, which use purification and
  swapping methods that lead to polynomial decay only.

\item {\it Entanglement percolation}. Recently, our
  collaborators, together with two of us, proposed using
  networks in which properties of the connectivity of the
  network enable the establishment of, and determine the
  probability of, entanglement on large distances. In
  \ourrefs\ we considered in particular {\it pure-state networks} on regular
  lattices, where the nearest-neighbor (NN) nodes share a
  non-maximally entangled pair of qubits, or more generally
  qudits (an entangled bond). We searched for local
  operations and classical communication (LOCC) protocols that
  lead to establishment of entanglement between remote nodes
  of the network.

\end{itemize}

At present, only a handful of studies that explicitly apply
percolation theory to problems in quantum information have
appeared in the literature. These articles, comprising
\ourrefs\ as well as applications to cluster states, are reviewed 
in Ref.~\onlinecite{kieling2007}.

Our results leading to the present work can be summarized as follows:
\begin{itemize}
\item For 1D chains we proved that even optimal LOCC
  strategies only allow the establishment of entanglement between
  distant nodes that decays exponentially with both the distance,
  as well as the quality of entanglement of NN bonds.
\item In 2D and higher dimensions, the possibilities for
  protocols are greatly expanded. The most straightforward,
  naive protocol--- the one we use as a baseline to evaluate
  other protocols--- is the one we term {\it classical
    entanglement percolation} (CEP), although it does
  involve obviously some quantum operations. This protocol
  begins with converting (using LOCC) each of the entangled
  bonds into a singlet ({\it i.e.} maximally entangled
  state) with probability $p$ (the so-called singlet
  conversion probability (SCP) \cite{V99,NV01}).
 After the conversion, each of the
  parties (nodes) knows obviously which of the bonds are now
  perfectly entangled.
 Using classical communications, the
  parties establish whether there exists an infinite 
  percolating cluster (or one spanning the lattice )
and who belongs to it. Then, by
  performing a series of entanglement swappings, it is
  possible to propagate entanglement between any two
  (widely) separated nodes that belong to the percolating
  cluster.  We call this scheme ``classical'' because it
  essentially maps the problem onto a classical bond
  percolation problem \cite{GrimmettA}, and its success or
  failure is equivalent to the success or failure of
  bond percolation in the same lattice. Namely, if $p>p_c$,
  where $p_c$ is the lattice-dependent critical percolation
  threshold (or, in other words, critical open-bond
  density), then entanglement between any two remote nodes
  can be established with probability $P>0$, which
  asymptotically does not depend on the distance.
\item At the same time that we introduced CEP, we 
  presented several schemes that went beyond the simple application
  of singlet conversion everywhere followed by entanglement swapping 
  along a path. We call these  {\it quantum entanglement
    percolation} (QEP) protocols because they use some kind of quantum pre-processing---
  for instance quantum
  measurements to transform one   percolation problem to another one.
  In these protocols
  CEP is also used, but is preceded by application of certain
  LOCC, which remove and replace bonds in the network lattice, resulting
  in a new lattice geometry with fundamentally different long-range
  properties.
  The  pre-processing  in QEP may greatly
  improve the possibility of entanglement percolation, either
  by reducing $p_c$, or by increasing $P$.
\end{itemize}

In this paper we present a more systematic and thorough
study of QEPs based on lattice transformations.  In
particular, we explore the entire parameter space with the
addition of Monte Carlo and series-expansion methods.
Although general principles remain to be found, we do discover
certain rules and regularities governing such strategies
(see \secr{sec:examples}). The paper is organized as
follows. In \secr{sec:models} we formulate and describe the
models and fix the notation.  Section
\ref{sec:nets_and_perc} discusses quantum networks and
percolation, whereas \ssecr{sec:transformations} deals with
LOCCs used for lattice transformations. Our main results are
presented in \secr{sec:examples}, which contains
descriptions of \Nexamp examples of transformations
enhancing CEP: i) the transformation of the kagom\'e lattice
to the square lattice, ii) the transformation of the
double-bond honeycomb (hexagonal) lattice to the triangular
lattice, iii) the transformation of the square lattice to
two decoupled copies of the square lattice, iv) the
transformation of the bowtie lattice to the decoupled
triangular and square lattices, and v) the transformation of
a triangular lattice with two different degrees of
entanglement into the decoupled square and triangular
lattices.  The cases ii), iii), and v) were discussed
already in Ref.~\onlinecite{perseguers:022308}, but we present
here more general and stronger results.  We conclude in
\secr{sec:conclusions}.

Over the past few decades, classical percolation theory has
seen the development of a number of quite sophisticated and
powerful methods, which we have adapted to the questions at hand.
In order to demonstrate the supremacy of QEP over CEP in
each of the cases i)-v) we employ these methods combined
with the methods of quantum information theory. An overview
of these methods is presented in the appendices:
\appendr{sec:majorization} presents the necessary facts from
majorization theory and singlet conversion protocols
\cite{NV01}, \appendr{sec:montecarlo} explains some details
of our Monte Carlo (MC) simulations, and finally
\appendr{sec:series} deals with series expansions.

\section{\label{sec:models} Formulation of the models}

\subsection{\label{sec:nets_and_perc} Quantum networks and percolation}

In this section, we describe the classes of quantum networks
and communication protocols that we investigated.

\subsubsection{ Networks of bipartite states \label{networksbipartite}}
 The networks we consider consist, prior to application of
communication protocols, of a collection of qubits
partitioned into pairs, each pair being prepared in an
identical pure state
$\ket{\alpha}\in\mathbb{C}^2\otimes\mathbb{C}^2$.  With
appropriate choice of bases, any such state can be written
\begin{equation}
    \ket{\alpha}=\sqrt{\alpha_0}\ket{00}+\sqrt{\alpha_1}\ket{11},
    \label{stateAB}
\end{equation}
where the Schmidt coefficients $\alpha_0,\alpha_1$ satisfy
$\alpha_0\geq\alpha_1$ and $\alpha_0+\alpha_1=1$. 
 We identify these pairs with bonds on a two-dimensional
lattice or edges on a graph, with two spatially-separated
qubits, one occupying each end of the bond. At regular
positions, a small set of these qubits are arranged near
enough to one another to allow measurements on any
subset. Such a set of qubits then constitutes a vertex (or
node or site) which is incident to each edge that
contributes a qubit to the vertex. This defines a lattice to
which we apply methods of statistical physics. In
particular, many results on the possibility of long-range
entanglement are described using percolation theory
\cite{GrimmettA,StaufferB}, and depend only on the graph
structure of the system.

\subsubsection{ Classical entanglement percolation }
 Given a lattice prepared as described above, we search for
the protocols consisting of local operations and classical
communication (LOCC) that yield the maximum probability of
achieving entanglement between two nodes separated by an
arbitrarily large distance.  The answer depends on the
single parameter $\alpha_1$ characterizing the state
$\ket{\alpha}$.  Even with a small palette of possible
operations, finding the globally optimal solution is a
difficult task.
Instead, we search for promising classes of protocols.
The simplest protocol consists of
attempting to convert the state associated with each bond to
a singlet via the ``Procrustean method'' of entanglement
concentration, which is the optimal strategy at the level
of a single bond
\cite{V99,NV01}. This conversion succeeds with probability
$p=2\alpha_1$ (See \appendr{sec:majorization}.), while
failure leaves the pair in a state with no entanglement. In
this way the system is described exactly by a bond
percolation process with open-bond density $p$. If there is
a path of open (maximally entangled) bonds connecting two
nodes, a sequence of entanglement swapping measurements, one
at each intermediate node, is then applied in order to entangle the first
and last node.  This protocol is the simplest example of
classical entanglement percolation.

\subsubsection{ Percolation theory for CEP. }

Here we review a few fundamental ideas in
percolation theory necessary to analyze CEP.  The nodes in
the lattice can be partitioned into sets such that each node
within a set is connected to each other node in the set via
a path of open bonds. Such a set of nodes is called an open
cluster, or sometimes simply a cluster.  The central fact of
percolation theory is that percolation processes on most
commonly-studied lattices in dimension $2$ and higher
exhibit a continuous phase transition as the bond density
passes through a critical value $p_c$.  For $p>p_c$ there
exists with probability one a unique (for the lattices we
study here) cluster of infinite mass (number of nodes),
while for $p<p_c$ all clusters are finite with probability
one. It follows that improving an LOCC protocol to obtain a
small change in $p$ can have a dramatic effect on the
probability of long-range entanglement.  In the
supercritical phase, long-range entanglement is possible,
while in the subcritical phase it is not possible.  Serving
as the order parameter is the density of the infinite
cluster $\theta(p)$ which we define via
\begin{equation}\label{thetadef}
  \theta(p)= P[A\in\iC],
\end{equation}
the probability that a fixed node $A$ (say, the node at
the origin) is
in the the infinite cluster $\iC$. When referring to
$\theta$ on a specific lattice, we use a symbol representing
the lattice as a subscript. The probability that two
selected nodes are members of the same cluster decays
roughly exponentially in their separation distance to
an asymptotic value of $\theta^2(p)$. (The length scale of decay is 
the correlation length $\xi(p)$.)
 This means, for the problem at hand, that in CEP
the probability that information can be propagated between
two nodes is asymptotically $\theta^2(p)$.

\subsection{\label{sec:transformations} Transformations of lattice structure}
It was proven in \ourrefs\ that CEP is not the optimal
strategy for establishing long-distance entanglement. The
demonstration is based on applying certain LOCC prior to
CEP. All of these pre-processing LOCC act on pairs of
qubits, and they either transform the state on a given bond,
or they replace two adjacent pure-state bonds by one, in
general, mixed state bond.

There are essentially three types of generalized measurement used: 

\begin{itemize} 
\item {\it Singlet conversion} The optimal LOCC singlet
protocol \cite{NV01} is used in three situations. In CEP
with single-bond lattices we apply it directly to the state
(1), which results in the singlet conversion probability
$p=2\alpha_1$. If the protocol is successful, the
 bond that is converted to the singlet can be used for
entanglement propagation (swapping), otherwise it is
useless. In CEP with a double-bond lattice we apply it
directly to the two copies of the state (1), which lives in
${\cal C}^4\otimes{\cal C}^4$, and has Schmidt coefficients
$\alpha_0,\sqrt{\alpha_0\alpha_1},\sqrt{\alpha_0\alpha_1},\alpha_1$,
{\it ergo} the singlet conversion probability is 
\begin{equation}\label{fourqubits}
p=\min\left\{1,2(1-\alpha_0^2)\right\}.
\end{equation}
 Finally, in all of the QEP strategies that we study, we
also apply singlet conversion to any remaining untouched
bonds after the lattice transformation.

\item {\it Entanglement swapping} This protocol~\cite{ZZHE93},
 illustrated in \figr{fig:swapping}a, consists of
performing the so--called Bell measurement on a pair of
qubits (b and c) in a node, i.e. a von Neumann measurement in
a basis of 4 maximally entangled orthonormal states (in the
computational basis). It allows conversion of a pair of
adjacent singlets into a singlet connecting end points (a
and d) with probability 1, i.e. allows for perfect
entanglement propagation in a connected cluster of
singlets. At the same time, when applied to a pair of
imperfect states (1), it produces a mixed state, which,
amazingly, has the average singlet conversion probability
equal to $p=2\alpha_1$. Unfortunately, this effect cannot be
iterated. When applied to the mixed states, entanglement
swapping reduces the singlet conversion probability.
This last point places a significant constraint on our
choice of lattice transformations.

We use entanglement swapping in both CEP and QEP, but the
two uses have very different effects. In the the case of
CEP, entanglement swapping is used to locally move
entanglement between neighboring nodes, an operation that is
repeated in hopes of transporting entanglement over long
distances (This is, roughly speaking, a brute-force method).
But with QEP, before attempting to transfer entanglement, we
search for a way to selectively apply entanglement swappings
to alter the geometry of the lattice and hence its
long-range connectivity properties as given by percolation
theory. The goal of this paper and future work is to
enumerate the rich possibilities and point a way towards a
general description of QEP.

\item{\it Worst case entanglement} Finally, in
\ourrefs\  the worst case
protocol was used, which maximizes minimal entanglement over
all measurement outcomes. This protocol consists also of
Bell measurement, but the basis is computational for one
qubit, and corresponding to eigenstates of $\sigma_x$ for
the second. When applied to qubits, it produces a mixed
state with the property that for all measurement outcomes,
the resulting pure states have the same singlet conversion
probability. We will not use this protocol here.
\end{itemize}

In order to describe the lattice transformations involved in
QEP, we need a more general formulation of the lattice than
the one given in \secr{networksbipartite} for CEP. It is
useful to define the percolation processes a bit more
precisely. We begin with a graph, that is, a set of edges
(bonds) $E$ and a set of vertices (nodes) $V$. We consider
embedding the graph in $\mathbb{R}^2$ in order
to treat geometric properties. In fact, the important
properties don't change if we force the vertices to occupy
points in $\mathbb{Z}^2$. The configuration of open and
closed bonds can be described by a probability sample space
$\Omega = \prod_{e\in E} \{0,1\}$, with points $\omega
=(\omega(e):e\in E)$, where $\omega(e)$ takes the values $0$
and $1$. We allow each bond to be open with a different
probability, {\it i.e.}  $\boldsymbol{p} = (p(e):e\in
E)$. The appropriate measure is a product measure on
$\Omega$ with marginal probabilities defined by
$\mu_e(\omega(e)=1)=p(e)$, $\mu_e(\omega(e)=0)=1-p(e)$. To
allow all transformations possible via LOCC, we take the
graph to be complete--- that is, all possible edges
$e=\langle v,w \rangle$ with $v,w\in V$ are present (In
fact, sometimes we need double bonds, as well.) For
percolation properties, any graph with fewer edges can be
identified with this complete graph by setting the
appropriate $p(e)$ to zero.

\begin{figure}
  \includegraphics[width=\widthtwo]{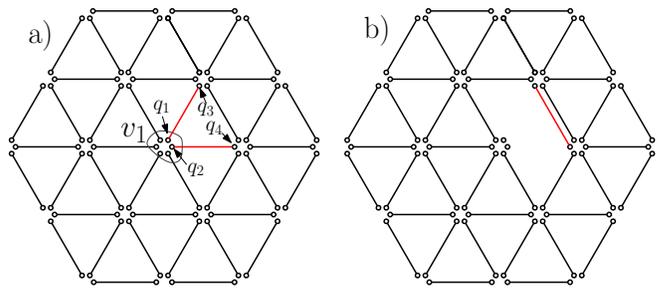}
  \caption{\label{fig:show_qep} (Color online) a) Mapping a quantum network
    to a percolation problem. Qubits are represented by
    small circles. Each vertex (node) contains six qubits.
    b) using entanglement swapping to transform the lattice
    structure. The two red (grey) bonds in a) are replaced by the red bond
  in b). This process can be continued to produce a double-bond
  hexagonal lattice.}
\end{figure}

Now we describe the correspondence of entanglement on the
physical system of qubits to this percolation formulation.
We denote the set of the indices of all qubits by $A$. Each
qubit $a\in A$ is assigned to a vertex $v=V(a)$, with, in
general, multiple qubits assigned to each vertex. For
example, a portion of a network that can be described by a
triangular lattice is shown in figure \figr{fig:show_qep}a,
where vertex $v_1$ contains the six circled qubits.  For
every pair of qubits $a,b\in A$ with $V(a)\ne V(b)$, we
denote the reduced state by $\rho_{ab}$, and by $S(a,b)$ the
singlet conversion probability (SCP), that is, the
probability of conversion of $\rho_{ab}$ to a singlet,
maximized over all possible measurements. Then we assign
$p(e) = S(a,b)$ for every $e=\langle V(a),V(b)\rangle$.
Initially, all reduced bipartite states have either $p(e)=0$
for separable states, or $p(e) = p = 2\alpha_1$ for the
prepared, pure, partially entangled states. In
\figr{fig:show_qep}a the edges with $p(e)=p$ are shown in
black, while edges with $p(e)=0$ are simply absent in the
diagram. Two bonds are colored red (grey) only to
show that they will be replaced by another bond (in the
sense of altering the SCP) via entanglement swapping. For
QEP, a successful preprocessing, entanglement swapping
operation such as that depicted in \figr{fig:swapping},
alters the percolation process by setting $p(\langle
V(a),V(b)\rangle)=0$ and $p(\langle V(c),V(d)\rangle)=0$ and
$p(\langle V(a),V(d)\rangle)=p$.  Likewise, in the example
in \figr{fig:show_qep}, an entanglement swapping measurement
on qubits $q_1$ and $q_2$ sets $p(\langle
V(q_1),V(q_3)\rangle)=0$, $p(\langle
V(q_2),V(q_4)\rangle)=0$, and $p(\langle
V(q_3),V(q_4)\rangle)=p$.  Each of the examples discussed
below conforms to this description.

\section{\label{sec:examples} Examples}

We consider \Nexamp examples of lattice transformations in
this paper, each exhibiting a different combination of
features, with implications for the analysis of protocols.
In particular, two transformations convert one lattice to
another, while the other three convert a lattice into two
decoupled lattices.  Four transformations involve lattices
with single bonds, while the fifth involves double bonds.
Quantities arising in the analysis, such as $\theta(p)$, are
analytic about $p=1$. In fact, the lowest-order term in the
expansion about $p=1$ is typically much larger than the
remaining terms even relatively far from $p=1$, with the
result that most of the interesting crossover behavior in
comparing protocols occurs for smaller values of $p$.  But
this behavior does not appear when using the distillation
procedure for double bonds that leads to (\ref{fourqubits}).
The reason is that distillation produces a saturation point
for $\theta(p)$ smaller than $p=1$.  This results in a much
stronger difference between classical and quantum protocols
in the high-density regime than does singlet conversion on
single bonds (two-qubit pure states). Four of the
transformations produce a smaller critical density on at
least one of the resulting lattices, which gives the most
pronounced advantage in the regime near the critical
density. The fifth example shows that on some lattices where
a particular QEP strategy is not advantageous, allowing
bonds of different strengths $p$ and $p'$ can produce
regions in the phase space $(p,p')$ where QEP is indeed
advantageous. Finally, consider comparing CEP in which
singlet conversion is applied to each bond separately, with
QEP that results in a single transformed lattice. In every
such case we find that QEP is better than CEP over the
entire (non-trivial) range of $p$. It is an open question
whether this is generic behavior.

We analyze the results of the transformation of each lattice
in three regimes: near the critical density (or densities);
near $p=1$; and between these two regimes. Arguments near the
critical densities typically rely simply on the fact that
long-range entanglement is impossible on a lattice with
density below the critical density. These results are the most
insensitive to details of the definitions of entanglement
and connectivity. Near $p=1$ we compute high-density
expansions for $\theta(p)$ and related quantities (see
\secr{sec:expansions}.)  Often the difference between the
performance of CEP and QEP in this regime is marginal. It is
important nonetheless to carry out the analysis if we hope
to make general statements about transformations that hold for all
$p$.  Between the critical regime and high-density regime
there are some techniques widely used in percolation theory
that may be useful, such as Russo's formula, which is used
to prove inequalities in the rate of change of
$\theta(p)$.  However, we leave these techniques for future
work and use Monte Carlo computations in the present paper.

\subsection{\label{sec:kag} Kagom\'e lattice to square lattice}

\begin{figure}
  \includegraphics[width=\widthone]{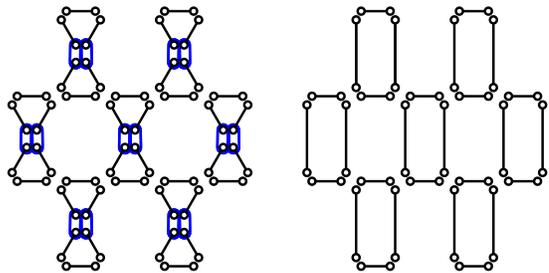}
      \caption{\label{fig:kagfive} One of three transformations of the kagom\'e lattice
	to the square lattice.  Pairs of qubits that are subjected
to entanglement swapping are marked with loops. }
\end{figure}

In our first example we compare CEP on the kagom\'e lattice to quantum
entanglement percolation consisting of transformation of the
kagom\'e lattice to the square lattice via entanglement
swapping at nodes specified in \figr{fig:kagfive}.
Although we do not treat them here, there are at least two
more ways to transform the kagom\'e lattice to the square
lattice using the same kind of entanglement swapping.
Rigorous bounds have been obtained for
$p_c(\text{kagom\'e})$ \cite{GrimmettA} (as
well as a high-precision Monte Carlo estimate
\cite{ziff1997}) while $p_c(\square)$ is known exactly,
proving that the transformation gives an advantage for $p$
lying between $p_c(\square)$ and the lower bound for
$p_c(\text{kagom\'e})$. As shown in \tabr{tab:series},
the series for $\theta_\square$ and $\thetakag$ are the same
to the first non-trivial order in $q=1-p$. But the next term
shows that $\theta_\square>\thetakag$ for $p$ close enough
to $1$. The MC data provides strong evidence that
$\theta_\square>\thetakag$ everywhere except near $p=1$,
where the statistical error is too large to distinguish the
curves. But as is evident from the lower plot in
\figr{fig:kagfive}, the terms in the expansions that we
computed dominate $\theta_\square-\thetakag$ already at
values of $p$ for which the MC data is still accurate.
Thus we find that QEP is advantageous over the entire range
of $p$.

\begin{figure}
  \includegraphics[width=\widthone]{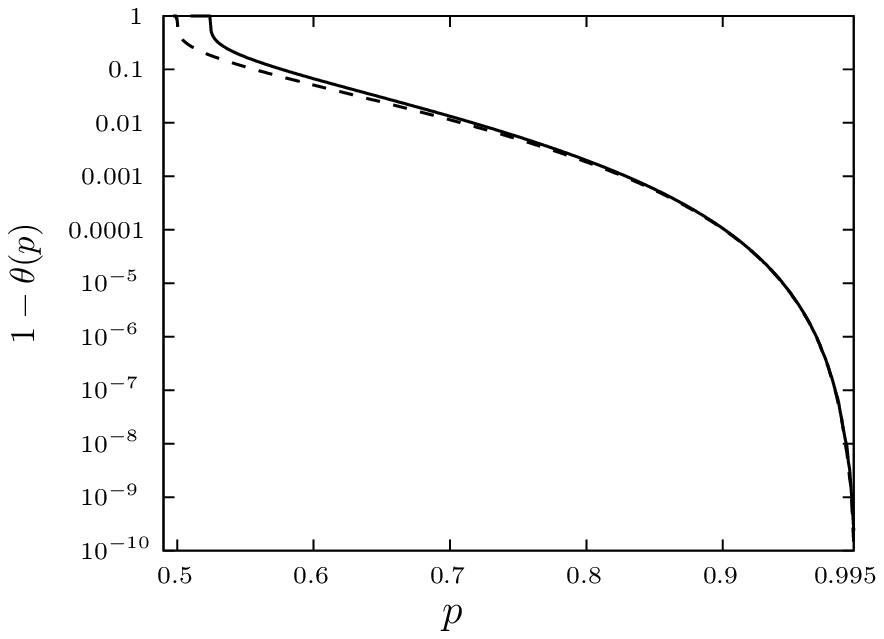}
  \vskip 10pt
  \includegraphics[width=\widthone]{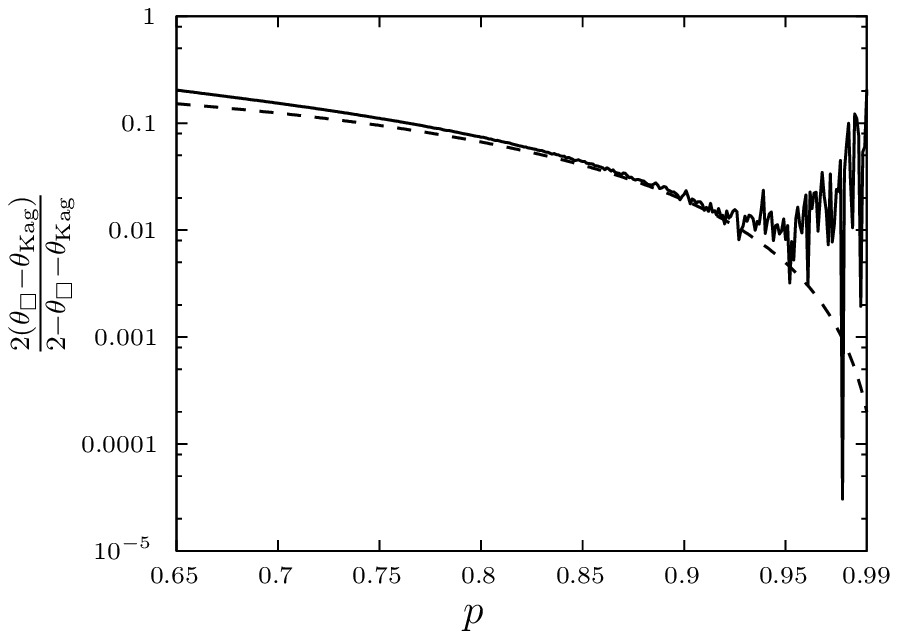}
      \caption{Upper plot: Monte Carlo of $1-\theta(p)$ {\it
       v.s.} $p=2\alpha_1$.  The solid line is on the square
       lattice. The dashed line is on the kagom\'e lattice.
       Lower plot: Normalized difference between
       $\theta_\square$ and $\thetakag$. The solid line is
       computed from MC data. The dashed line is computed from
       expansions from \tabr{tab:series}.  }
      \label{fig:sq_and_kag}
\end{figure}

\subsection{Double-bond hexagonal lattice to triangular lattice}

\begin{figure}
  \includegraphics[width=\widthone]{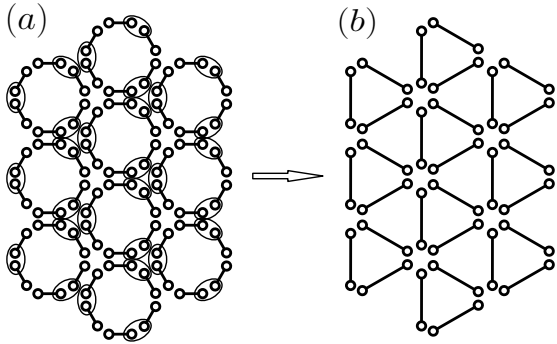}
      \caption{\label{fig:hontrtwo} (a) Double-bond
   hexagonal lattice. (b) transformed into the triangular
   lattice (b). Pairs of qubits that are subjected to
   entanglement swapping are marked with loops. }
\end{figure}

The hexagonal lattice with double bonds, each in the state
specified by (\ref{stateAB}) (see \figr{fig:hontrtwo}a), was
discussed in \ourrefs, in which it was shown that a
transformation of the lattice to a triangular lattice offers
an advantage over CEP protocols for values of $p$ between
$p_c(\triangle)$ and another critical value defined below.
  Here we extend the analysis to determine over the entire
range of $p$ which of three protocols gives the highest probability of
long-range entanglement. The first of the three protocols, which
we call CEP I, consists of performing an optimum singlet
conversion to each bond separately so that the probability
of getting at least one singlet connecting two nodes is
$p'=1-(1-p)^2$. Communication on the resulting lattice is
then determined by bond percolation on the hexagonal lattice
with bond density $p'$. In particular, the critical density
(see \tabr{tab:pcs}) is $p=p_c(\text{CEP I})\approx 0.4107$.
In the second protocol, we perform a more efficient
conversion, namely distillation, on all four qubits in a
double bond. This protocol succeeds in producing a maximally
entangled pair with probability
$p''=\min\{1,2(1-\alpha_0^2)\}$ (See
 (\ref{fourqubits}) and \appendr{sec:majorization}.)  This results in bond
percolation on the hexagonal lattice with density
\begin{equation}\label{distillp}
p''=\min \left\{1,2\left[1-\left(1-\frac{p}{2}\right)^2\right]\right\}.
\end{equation}
  We refer to this method as CEP II, with critical density
$p=p_c(\text{CEP II})\approx 0.358$.  In the third method,
the entanglement swapping that maximizes SCP is applied at
every other node (see \figr{fig:hontrtwo}), converting
the double-bond hexagonal lattice to the triangular lattice
with bond density $p$, with $p_c(\triangle)\approx 0.347$.

\begin{figure}
  \includegraphics[width=\widthone]{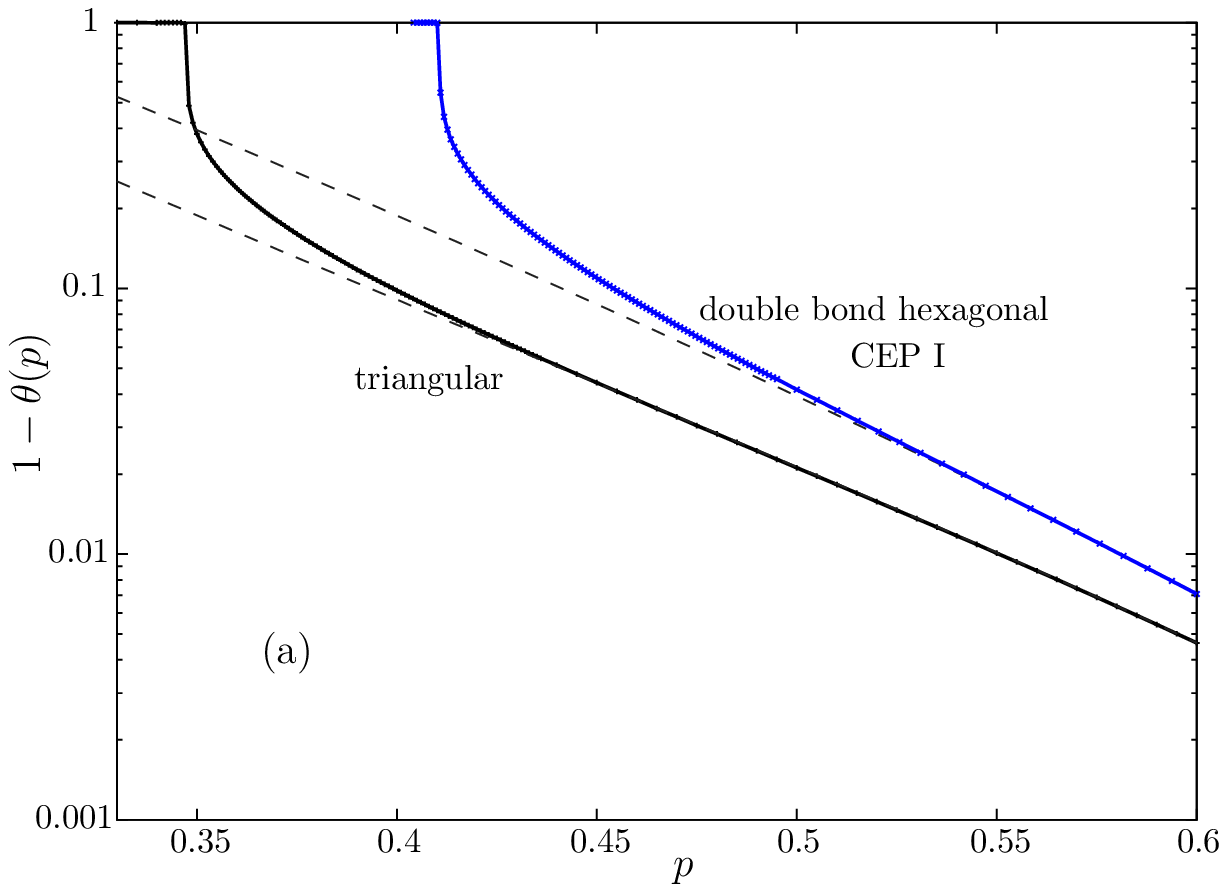}
  \vskip 10pt
  \includegraphics[width=\widthone]{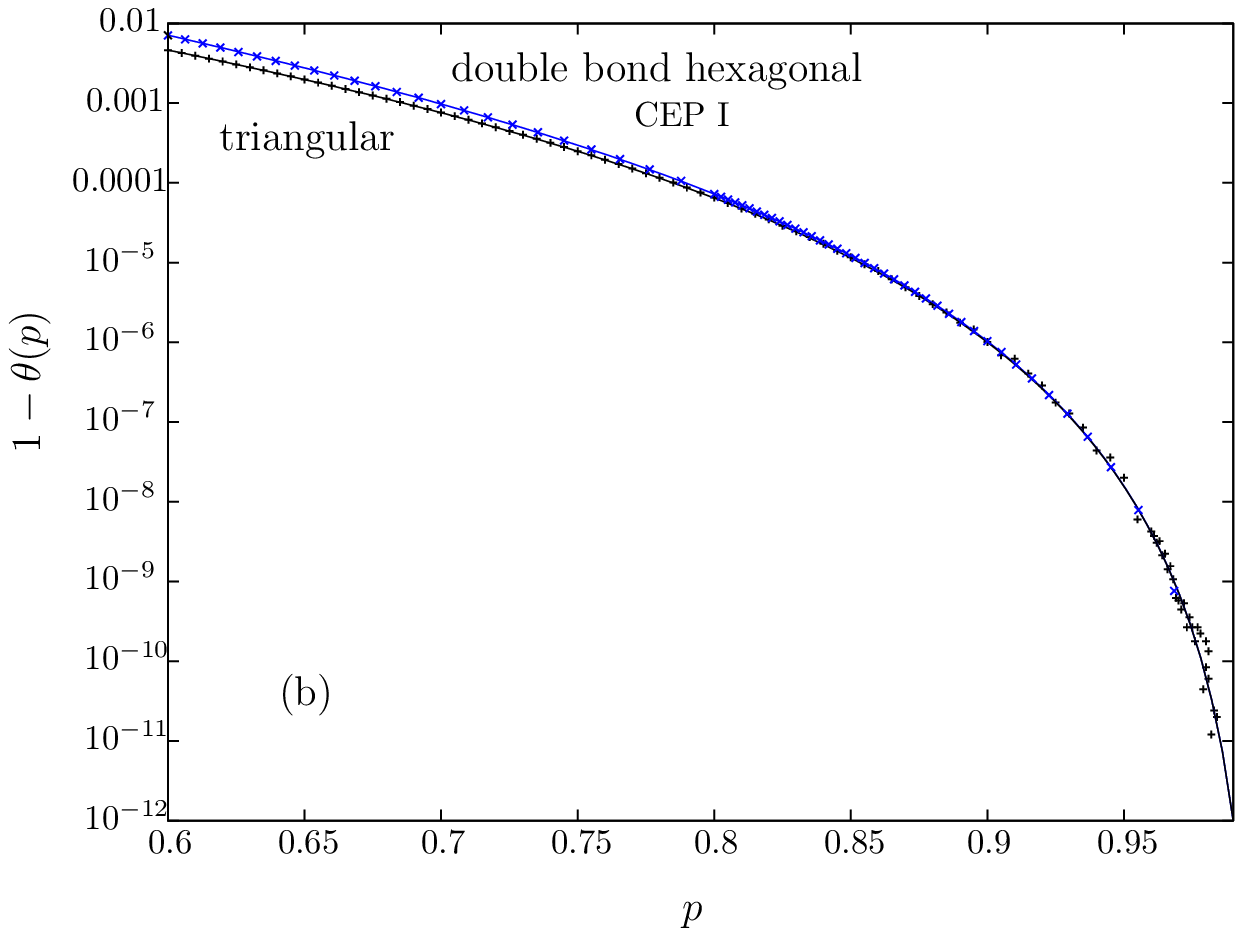}
  \vskip 10pt
  \includegraphics[width=\widthone]{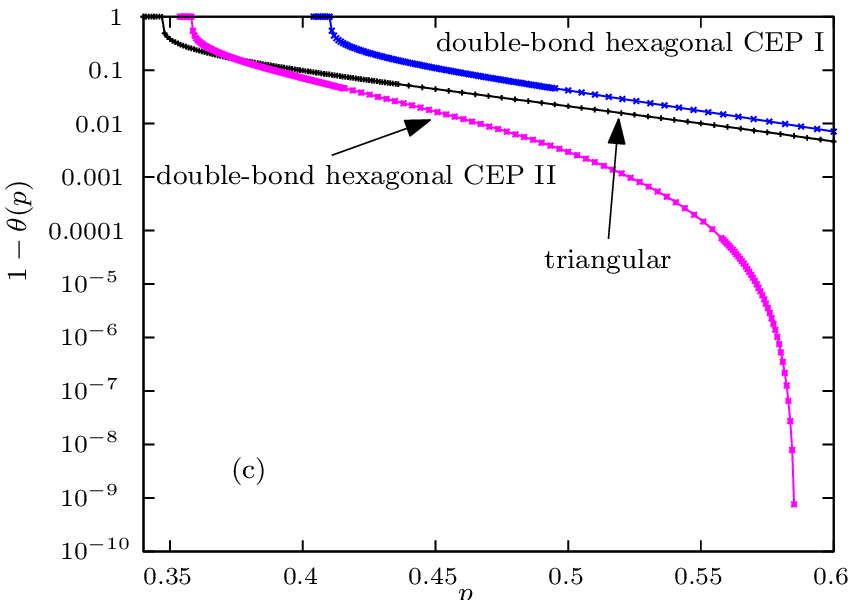}
      \caption{
    Monte Carlo and series expansions about $p=1$ of $1-\theta(p)$
   {\it v.s.} $p=2\alpha_1$     for various strategies on the
   double-bond hexagonal lattice.
  (a) Solid lines are Monte Carlo data. Dashed lines are series expansions.
  (b) Solid lines are MC. Series expansions are indistinguishable from MC
   over the entire plot.
  (c) Monte Carlo data.
      \label{fig:trihexone}}
\end{figure}
The three protocols are compared in \figr{fig:trihexone}.
We first observe that each method fails below its
corresponding critical density, and these are known exactly.
To compare the methods away from the critical points, we
computed series expansions of $\theta(p)$ about $q=0$ with
results listed in \tabr{tab:series}.  Defining
$\tilde\theta(q)=\theta(1-q)$ we see that
\begin{equation}\label{thetacepi}
\tilde\theta_\text{CEP I}(q) = \tilde\theta_{\hexagon}(q^2) = 1-q^6-3q^8 +\ldots
\end{equation}
Comparing (\ref{thetacepi}) to $\theta_\triangle(p)$ from
\tabr{tab:series} we see that for $p\to 1$ the conversion to
the triangular lattice is better than CEP~I. As is evident
from Figs. \ref{fig:trihexone}a and \ref{fig:trihexone}b, the
two curves have the same leading behavior as $p\to1$, with
the result that the the Monte Carlo cannot distinguish them
in this region.  However, the series expansion lies well
within the statistical error of the Monte Carlo points, even
relatively far from $p=1$, so we can be confident that
$\theta_\triangle > \theta_\text{CEP I}$ even in the region
where the MC cannot distinguish the curves. Finally, the MC
clearly shows that $\theta_\triangle > \theta_\text{CEP I}$
for smaller $p$, where the expansion fails.  Turning now to
CEP II, we know that $\theta_\triangle > \theta_\text{CEP
II}$ for $p_c(\triangle) < p < p_c(\text{CEP II})$ and from
(\ref{distillp}) that $\theta_\text{CEP II}=1$ for $p\ge
2-\sqrt{2}$. It follows that there must be a crossover
point. The MC data suggests that this occurs for $p\approx
0.375$. In summary, we see that QEP gives an advantage over CEP I for
all $p$, but that CEP II, which is more efficient in its use
of double bonds, is better than QEP at high densities.

\subsection{\label{sec:doubsquare} Square lattice to two decoupled copies of the square lattice}

\begin{figure}
  \includegraphics[width=1.0\columnwidth]{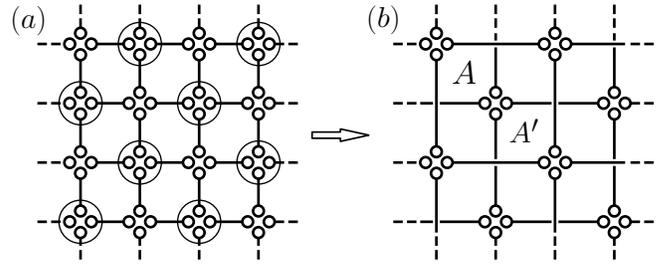}
  \caption{Doubling the square lattice. Large circles represent
 nodes at which entanglement swapping is performed.
   }
      \label{fig:doubsq}
\end{figure}

This transformation replaces the square lattice with
two decoupled copies of the square lattice. To effect
the transformation,
at selected nodes the two horizontally opposing bonds are joined
into one bond and likewise with the vertically opposing bonds. This
procedure is applied at every other node, while staggering
by one node when shifting rows as shown in \figr{fig:doubsq}.
\begin{figure}
  \includegraphics[width=\widthone]{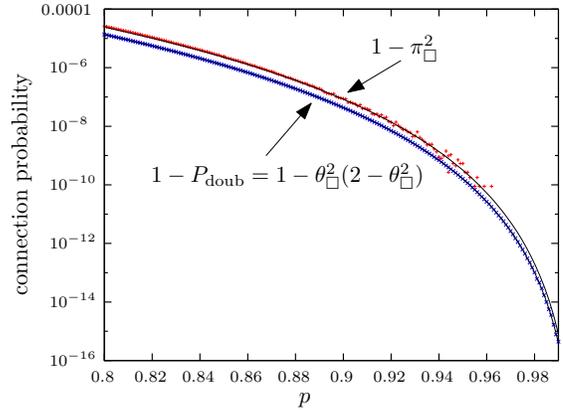}
  \caption{ Monte Carlo and series expansions for doubling the
  square lattice.
 Lower points: $1-P_\text{doub}$ {\it v.s.} $p$.
  Upper  points: $1-\pisq^2$  {\it v.s.} $p$. Solid lines
  are series expansions.
   }
      \label{fig:pi_pdoub_asymp}
\end{figure}
\begin{figure}
  \includegraphics[width=\widthone]{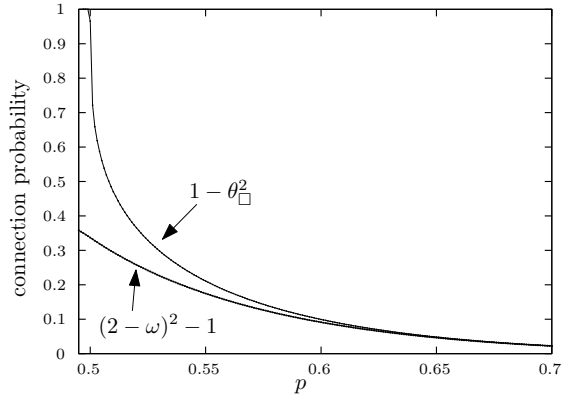}
  \caption{ Monte Carlo for doubling the
  square lattice. Upper points: $1-\theta_\square^2$ {\it v.s.}
   $p$. Lower points: $(2-\omega)^2 - 1$ {\it v.s.} $p$.
   }
  \label{fig:theta_omega_nolog}
\end{figure}
Because the transformation splits the original lattice into
two disjoint lattices, each taking half the surviving nodes,
we cannot compare connection between a single pair of nodes
before and after the transformation. Rather, we consider on
the original lattice connection between, on the one hand, a
pair of nodes $A,A'$ separated by a small distance, and on
the other hand, the same pair translated a distance much larger
than the correlation length to a pair $B,B'$.  We choose $A$
and $A'$ so that each one goes to a different lattice in the
transformation. After the transformation, there is the
possibility of connection between $A$ and $B$ on one lattice
and between $A'$ and $B'$ on the other. We choose the nodes
as shown in \figr{fig:doubsq}. In addition to the
complication of splitting, this example is subtle because
the critical density is the same on all three lattices.  In
Ref.~\onlinecite{perseguers:022308} it was shown that just
above $p_c(\square)=1/2$, the transformation gives an
advantage over CEP as measured by the choice of nodes
$A,A'$.

Here we examine the connectivity for all $p>p_c(\square)$.  The
probability that at least one of the pairs on the decoupled
lattices is connected is
\begin{equation*}
 P_{\text{double}}= 1-(1-\theta_\square^2)^2 = \theta_\square^2(2-\theta_\square^2).
\end{equation*}
We compare $P_{\text{double}}$ to the probability on the original lattice
that at least one of $A$ and $A'$ is connected to at least one
of $B$ and $B'$. This probability is $\pi_\square^2$ where
\begin{equation}\label{pisq}
\pisq=P[A\in\iC \text{ or } A'\in\iC].
\end{equation}
 (In this paper `or' in `$X$ or $Y$' does {\it not } mean exclusive or.)
We prove that the doubling is advantageous in the limit
$p\to 1$ using the power series about $q=0$ for
$\theta_\square$ and $\pi_\square$ given in
\tabr{tab:series}. This follows from comparing
\begin{equation*}
 \pi_\square^2 = 1 - 8q^8 - 36q^{10}+ \ldots
\end{equation*}
and
\begin{equation*}
 \theta_\square^2(2-\theta_\square^2) = 1 - 4q^8 - 32q^{10}+ \ldots.
\end{equation*}
Moreover, Monte Carlo evidence suggests that the
transformation improves on CEP for all $p>p_c(\square).$ In
\figr{fig:pi_pdoub_asymp} one sees from the accuracy of the
series expansion in the region of high-quality MC data that
that the advantage is maintained even when the Monte Carlo
becomes noisy near $p=1$.  The same MC data is not useful
near $p_c(\square)$ for two reasons. Firstly, both
$P_\text{doub}$ and $\pi_\square^2$ vanish at the same
critical point with infinite slope. Secondly, the systematic
error due to finite lattice size and large fluctuations in
cluster statistics further complicate distinguishing the
curves. We instead make use of an alternate expression for
$\pi$, that is $\pi_\square=\theta_\square(2-\omega)$ where
$\omega=P[A\in\iC|A'\in\iC]$.  The condition for advantage
over CEP then becomes
\begin{equation}\label{alternate}
(2-\omega)^2 < 2-\theta_\square^2.
\end{equation}
The MC data for $\omega$ was generated by considering the
largest cluster in the finite lattice to represent the
infinite cluster $\iC$ even if it is not a spanning cluster.
The MC data supports (\ref{alternate}), and furthermore, shows
no evidence of non-analyticity at $p_c(\square)$, as is
evident in \figr{fig:theta_omega_nolog}.

The foregoing analysis of the doubled square lattice was
based on the choice of $A,A'$ shown in \figr{fig:doubsq}.
But this is not the only reasonable choice. In fact the
question of whether the doubling transformation is better
than CEP depends somewhat on the details of how the nodes
are chosen. Although this is an extreme example, similar
questions arise in analyzing other transformations. Thus,
the ambiguity in the measures comparing the
various protocols must be addressed in future work.

\subsection{Bowtie lattice to decoupled triangular and square lattices}
\begin{figure}
  \includegraphics[width=0.6\columnwidth]{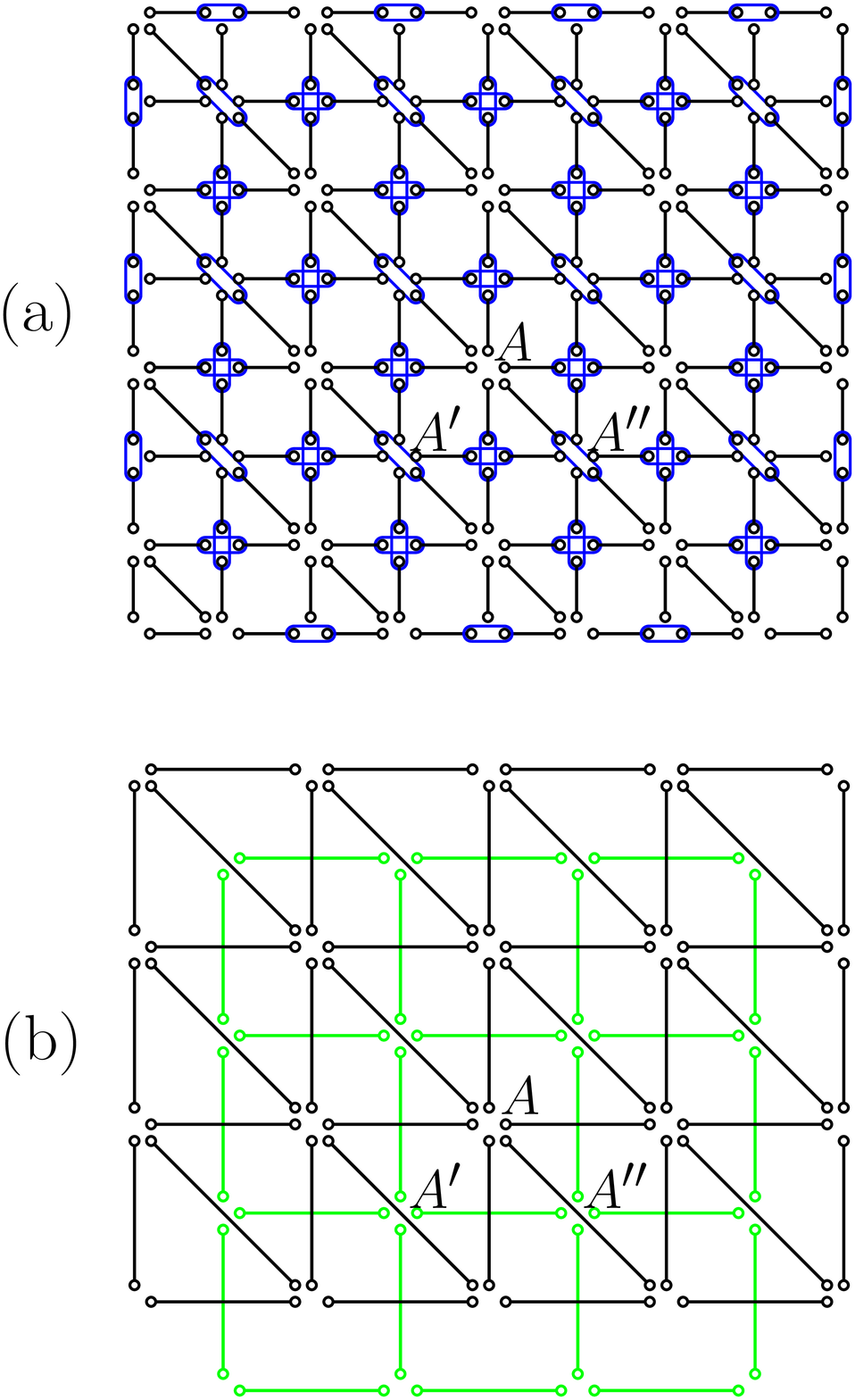}
      \caption{\label{fig:bowtie} Transformation of the bowtie lattice to
  decoupled square and triangular lattices. Loops marking pairs of qubits
 represent swapping measurements. }
\end{figure}

This is our second example that transforms a lattice into
two decoupled lattices. Figure \ref{fig:bowtie} shows the
measurements that decouple the bowtie lattice into the
square lattice and the triangular lattice.  For $p$
satisfying $p_c(\triangle)<p<p_c(\bowtie)$ (the exact critical
values are listed in \tabr{tab:pcs}), the transformation is obviously
advantageous. 

In carrying out further analysis, the bowtie lattice
presents a complication not present in other lattices
studied in this paper: While the other lattices are regular
in the sense that each vertex has the same environment up to
rotations and reflections, the bowtie lattice has two kinds
of vertices in this sense.  Notice first that all the
surviving nodes on the square and triangular lattices shown
in \figr{fig:bowtie} were
generated from nodes of coordination number $z=6$ on the
bowtie lattice, with all nodes of $z=4$ disappearing. Each
node of $z=6$ on the bowtie lattice has four nearest
neighbors of $z=6$, discounting nodes of $z=4$ (for
example node $A$ in \figr{fig:bowtie}.) Furthermore, two of
these nearest neighbors are connected by a diagonal bond
(node $A''$) and two are not (node $A'$.)

\begin{figure}
  \includegraphics[width=0.85\columnwidth]{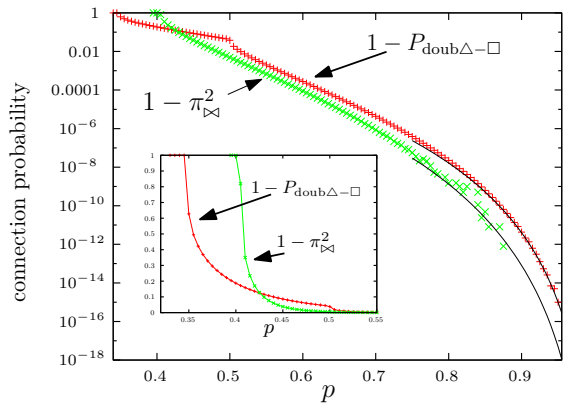}
      \caption{\label{fig:bt_sq_tr_log}  Monte Carlo estimates of
  $\pibt$ and $P_{\text{doub}\triangle-\square}$. The inset
 is the same data on a linear scale. Solid lines are high-density
 expansions.
  }
\end{figure}

In treating the example of doubling the square lattice in
\secr{sec:doubsquare}, the fact that the lattice
decouples forced us to consider connections between two
widely separated pairs of nodes.  We treat the present
example in the same way, except that the more complicated
local structure of the bowtie lattice forces us to consider
a cluster of three nodes $A,A',A''$ rather than a pair.  In
\figr{fig:bowtie}b we see that $A',A''$ are sent to the
square lattice and $A$ is sent to the triangular lattice. We
consider a distant cluster of nodes $B,B',B''$ related to
$A,A',A''$ by translation, and examine the probability that a
connection exists between at least one of $A,A',A''$ and at
least one of $B,B',B''$. On the bowtie lattice, this
probability is $\pibt^2$ where
\begin{equation}\label{pibt}
\pibt = P_{\bowtie}[A\in\iC\text{ or } A'\in\iC\text{ or }
A''\in\iC].
\end{equation}
If instead we decouple the lattices via the lattice transformation, this
probability is given by
\begin{equation*}
  P_{\text{doub}\triangle-\square} =  \theta_\triangle^2 +
    \pisqb^2 - \theta_\triangle^2\pisqb^2,
\end{equation*}
where 
\begin{equation}\label{pisqb}
\pisqb = P_\square[A'\in\iC \text{ or } A''\in\iC].
\end{equation}
A comparison of Monte Carlo estimates and high-density
expansions of these quantities is shown in
\figr{fig:bt_sq_tr_log}.
The MC data shows that for small densities, the lattice transformation
gives an advantage, while for higher densities CEP is the better protocol.
The cross-over occurs at $p\approx 0.425$. As with the other examples, the
curves representing the series expansions suggests that no cross-over
occurs in the high-density region.

\begin{table}[]
\begin{center}
    \begin{tabular}{@{\extracolsep{10pt}} l l }
    Lattice & $p_c$ for bond percolation \\
    \hline      
    triangular & $2\sin(\pi/18)\approx 0.347$\\
    bowtie & $\approx 0.4045$\\
    square     & $0.5$\\
    kagom\'e   & $\approx 0.5244053$ \quad {\text MC estimate } \\
    hexagonal  & $1-2\sin(\pi/18)\approx 0.653$\\
    \end{tabular}
\end{center}
\caption{\label{tab:pcs} $p_c$ for bond percolation on some lattices.
  $p_c(\bowtie)$ is the unique root of $1-p-6p^2+6p^3-p^5$. All critical
  densities are exact\cite{GrimmettA} except for $p_c(\text{kagom\'e})$\cite{ziff1997}. }
\end{table}

\begin{table}[]
    \begin{tabular}{@{\extracolsep{10pt}}  l l }
      $\theta_\square(p)$ & $1 -q^4 -4q^6$ \\
      $\pisq(p)$    &  $1-4q^{8} -18q^{10}$\\
      $\pisqb(p)$    &  $1-q^{6} +q^7 -8q^8$ \\
      $\theta_\triangle(p)$ & $1 -q^6 -6q^{10}+6q^{11}-6q^{12} -21q^{14}+42q^{15}$ \\
      $\theta_{\hexagon}(p)$ &  $1-q^3-3q^4-6q^5-25q^6$ \\
      $\thetakag(p)$ & $1 -q^4 -6q^6$ \\
      $\pibt$ & $1-4q^{14}$ \\
    \end{tabular}
\caption{\label{tab:series}A few terms in series expansions
about $q=0$ of $\pi$ and $\theta$ for various lattices.
$\theta_\square, \theta_\triangle, \theta_{\hexagon},
  \thetakag$ are defined via (\ref{thetadef}).
 $\pisqb$ is defined via (\ref{pisqb}),  $\pisq$ via (\ref{pisq}) ,
and $\pibt$ via (\ref{pibt}).
}
\end{table}

\subsection{Asymmetric triangular lattice}

\begin{figure}
  \includegraphics[width=0.7\columnwidth]{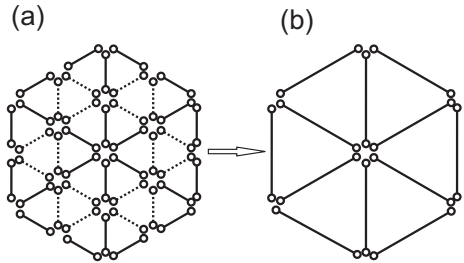}
      \caption{\label{fig:asymtr} a) The asymmetric
   triangular lattice. Solid bonds represent the state
   $\ket{\alpha}$ (density $p$). Dashed bonds represent the
   state $\ket{\alpha'}$ ( density $p'$). b) Entanglement
   swapping is performed on pairs of solid bonds resulting
   in a triangular lattice with bond density $p$. The
   remaining dashed bonds form a kagom\'e lattice which is
   then transformed into a square lattice with bond density
   $p'$ as in \ssecr{sec:kag}.  }
\end{figure}
With this example we demonstrate that QEP can succeed when
the initial bonds are not all in the same state, but fail if
they are in the same state.  Consider the lattice in
\figr{fig:asymtr}a composed of two different kinds of bonds,
each of the form given by (\ref{stateAB}).  The solid and
dashed bonds represent the states $\ket{\alpha}$ (with
$p=2\alpha_1$) and $\ket{\alpha'}$ (with $p'=2\alpha_1'$)
respectively.  In general, $p\ne p'$. We consider two
entanglement distribution protocols: a) classical
entanglement percolation on the original lattice and b) QEP
consisting of creating two decoupled lattices (the square
and triangular lattices) via entanglement swapping, followed
by CEP on each of the resulting lattices. The transformation
is described in \figr{fig:asymtr}. Note that, for $p'=0$,
the initial lattice is a triangular lattice with serial
double bonds and therefore the critical point
$p=\sqrt{p_c(\triangle)}\approx 0.589$. On the other hand for $p=0$
and $p'\ne 0$  the
initial lattice is the kagom\'e lattice.

\begin{figure}
  \includegraphics[width=0.8\columnwidth]{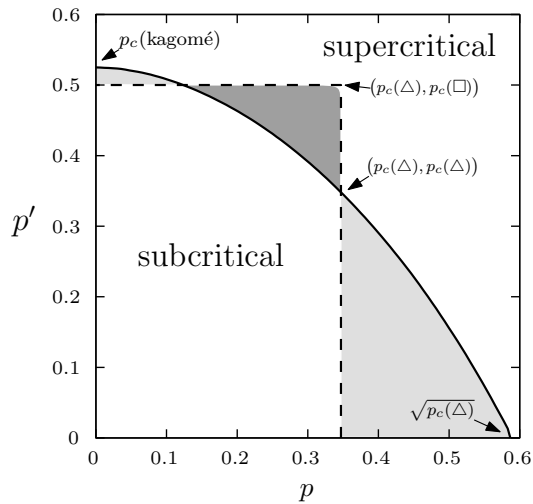}
      \caption{\label{fig:asymtr_phase} Phase diagram for
    the asymmetric triangular lattice. The heavy solid curve
    (obtained by Monte Carlo) separates the supercritical
    and subcritical regions on the original asymmetric
    triangular lattice (CEP). The dashed curve composed of
    two line segments separates the supercritical and
    subcritical regions after the transformation
    (QEP). After transformation, the supercritical region is
    defined to be the region for which at least one of the
    disjoint lattices is supercritical. In the light gray
    regions, long-range entanglement is possible with QEP,
    but not with CEP.  In the dark gray region it is possible with
    CEP, but not with QEP.}
\end{figure}

We first examine the most robust measure of a protocol's
effectiveness--- the binary measure that tells whether 
long-range entanglement is possible or not. The phase diagrams
for this example before and after the transformation are
shown in \figr{fig:asymtr_phase}. We see that there are
regions in the phase space for which QEP is better than CEP,
and vice versa.

In the region in which both protocols allow long-range
entanglement, a more detailed measure similar to those in
previous sections is necessary. For example, for $p=p'$, we
performed an analysis similar to the one used for doubling
the square lattice. The two critical boundaries intersect at
$p=p'=p_c(\triangle)$, but comparing connectivity just above
the critical point shows that the transformation does not
improve the probability of long-range entanglement. Likewise
for $p=p'$ and $p$ near $1$, series expansions showed that
the transformation is not an improvement on CEP. We have not
yet determined whether  QEP in this scenario succeeds for some
other $p=p'$, but this seems unlikely.

\section{\label{sec:conclusions} Conclusions}

In this paper we have considered the problem of establishing
long-distance entanglement in quantum pure-state networks on
regular 2D lattices. We have discussed in detail several
examples of quantum entanglement percolation strategies that
are better than the corresponding classical strategies,
i.e. those consisting of a direct attempt to convert bonds
into singlets. Our results illustrate nicely the interplay
between quantum information theory and classical percolation
theory. Despite the fact that we do find certain rules and
regularities governing QEP strategies, many questions remain
open. For instance, we cannot say anything about optimality
of our QEP protocols--- and most probably they are not
optimal. All of our protocols involve LOCC acting on pairs
of qubits only; can multipartite LOCC, which leads
inevitably to creation of multipartite entanglement, help?
We know that for sufficiently large initial entanglement,
perfect entanglement between remote nodes may be established
with distance-independent probability $P>0$. Is it also
possible for arbitrarily small initial entanglement?

\section*{Acknowledgements}
 We thank A. Ac\'in, J.I. Cirac, and D. Cavalcanti for
discussions. We thank S. Perseguers for kindly supplying
figures. We acknowledge support of the EU IP Programme
``SCALA'', and Spanish MEC grants (FIS 2005-04627,
Consolider Ingenio 2010 ``QOIT''). J.W. thanks Lluis Torner
and ICFO for hospitality.  He was partially supported by NSF
grant DMS 0623941. M. L. thanks also the Alexander von Humboldt
Foundation for support.

\appendix

\section{\label{sec:majorization} Majorization Theory }

Majorization theory has been applied to questions of
transforming one bipartite pure state to another by LOCC. In
particular, a theorem due to Vidal\cite{NV01} gives the
probability that such a transformation can be achieved via
an optimal protocol (without specifying that protocol.) We
state the result and apply it to the distillation procedure
used in this paper. We must first introduce a certain
partial order on vectors. Consider two real, $d$-dimensional
vectors $\mvec{r}$ and $\mvec{s}$.  We define the vector
$\mvec{r^\uparrow}$ by reordering the elements of $\mvec{r}$ into
non-decreasing order, and likewise with $\mvec{s}$.  We say
that $\mvec{r}$ is submajorized by $\mvec{s}$, denoted by
$\mvec{r}\subm\mvec{s}$, if
\begin{equation}\label{submaj}
 \sum_{i=0}^k r_i^\uparrow \ge  \sum_{i=0}^k s_i^\uparrow,
\end{equation}
for all $k=0,\ldots,d-1$. Denoting the vector of Schmidt
coefficients of $\psi$ by $\mvec{\lambda}(\psi)$, the theorem
states that $\ket{\psi}$ can be transformed into $\ket{\phi}$
with probability $p$, where $p$ is the largest number on $[0,1]$
such that $\mvec{\lambda}(\psi)\subm p \mvec{\lambda}(\phi)$.

A simple, relevant application is the computation of the
probability that optimal conversion of a state
$\ket{\psi}\in\mathbb{C}^d\otimes\mathbb{C}^d$ to a singlet
will succeed. Application of (\ref{submaj}) gives
$p=\min\{1,2(1-\alpha_0)\}$.

Now we consider distillation, defined here as the optimal
protocol for converting $n$ pure states $\ket{\alpha_i}\in
\mathbb{C}^2\otimes\mathbb{C}^2$ to $n-1$ product states and
one maximally entangled state in the Schmidt bases.  This
operation is useful, for instance, in attempting to get a
single, maximally-entangled bond from $n$ bonds connecting
two nodes.  Explicitly,
\begin{equation*}
 \ket{\alpha_i} = \sqrt{\alpha_{i,0}} \ket{00}_i +\sqrt{\alpha_{i,1}} \ket{11}_i, 
\end{equation*}
with $\alpha_{i,0}\ge\alpha_{i,1}$ and elements of the
ordered set $(\alpha_{0,0},\alpha_{1,0}, \ldots,
\alpha_{n-1,0})$ non-increasing. As usual, $\ket{jk}_i$ is
shorthand for $\ket{j}_{i,0}\otimes\ket{k}_{i,1}$.  For
$\mvec{j}\in\{0,1\}^n$ define the bijective numeration
$k(\mvec{j})=\sum_{i=0}^{n-1} 10^i j_i$ and
$\gamma_k=\gamma_{k}(\mvec{\alpha})=
\alpha_{0,j_0}\alpha_{1,j_1}\cdots\alpha_{n-1,j_{n-1}}$.
Then we can write
\begin{equation}\label{refactor}
\bigotimes_{i=0}^{n-1} \ket{\alpha_i} 
    = \sum_{\mvec{j}: k(\mvec{j})=0}^{2^n-1} \sqrt{\gamma_k}
     \bigotimes_{m=0}^{n-1} \ket{j_m j_m}_m.
\end{equation}
In order to apply the theorem, we need to show that this
state can be written as a bipartite state with Schmidt coefficients
that can be chosen to satisfy our distillation condition.  To this end
collect all the first qubits of the bipartite states and all
the second qubits, defining
\begin{equation*}
\ket{k(\mvec{j})}_a = \bigotimes_{i=0}^{n-1} \ket{j_i}_{i,0} \ \text{ and }
\ket{k(\mvec{j})}_b = \bigotimes_{i=0}^{n-1} \ket{j_i}_{i,1}, 
\end{equation*}
so that (\ref{refactor}) becomes
\begin{equation}\label{bigbipartite}
\bigotimes_{i=0}^{n-1} \ket{\alpha_i} 
    = \sum_{\mvec{j}: k(\mvec{j})=0}^{2^n-1} \sqrt{\gamma_k}
   \ket{k(\mvec{j})}_a \ket{k(\mvec{j})}_b.
\end{equation}
Note that $\sum_{k=0}^{2^n-1} \gamma_k = \prod_{i=0}^{n-1} (\alpha_{i,0}
  + \alpha_{i,1}) = 1^n = 1$ 
so that (\ref{bigbipartite}) defines a state in $\mathbb{C}^{2n}\otimes\mathbb{C}^{2n}$
already in Schmidt form.  The submajorization condition is
\begin{equation*}
 (\gamma_{2^n-1},\gamma_{2^n-2},\ldots,\gamma_{0}) \subm
  \left(0,0,\ldots,\frac{p}{2},\frac{p}{2}\right),
\end{equation*}
for which the only nontrivial inequality is the penultimate one
$1-\gamma_{0} \ge p/2$. Thus the maximum distillation probability
is
\begin{equation*}
 p = \min \left\{ 1, 2(1-\alpha_{0,0}\alpha_{1,0}\cdots \alpha_{n-1,0})\right\}.
\end{equation*}
Note that this result agrees with the special case in (\ref{fourqubits}).

\section{\label{sec:montecarlo} Monte Carlo estimates of $\theta(p)$ and $\pi(p)$ }

We computed Monte Carlo estimates of $\theta(p)$ and
$\pi(p)$ using the Hoshen-Kopelmann \cite{hoshen1976}
algorithm with modifications for efficiency
\cite{nakanishi1980} and the Mersenne twister
\cite{nishimura2000} random number generator. Together with
series expansions and exactly known critical densities, the
quality of the data we obtained is more than sufficient to
determine which transformations are advantageous. We
estimated $\theta(p)$ by computing the mean density of
the largest cluster on an $L\times L$ lattice.
Near the critical density we typically used lattices of size
$L=2$--$5 \times 10^5$. We took the value of $L$ at the inflection point of
plots of $\theta(p)$ {\it v.s.} $L$ at fixed $p$ as
estimates of the correlation length $\xi(p)$. We typically
found that $\xi(p)>L$ for $p-p_c < 0.0005$. We only expect
significant systematic error in this region, but this poses
no problem because the curves are never close to one another
in these regions. (We reformulated the problem when this is
the case.) We computed statistical error, but the
error bars are at most barely visible on the plots, so we
omitted them. Exceptions are very near $p=p_c$ where
fluctuations in the size of the largest cluster become
large, and near $p=1$, where finite clusters are rare, so
collecting sufficient samples to distinguish curves is
too expensive. We discuss the effects of these errors in the
main body of the paper.

We wrote a single computer code to study all the lattices.
The code supports lattices with vertices that occupy points
on $\mathbb{Z}^2$ with bonds connecting each pair of nearest
neighbors as well as a diagonal bond from $(i,j)$ to
$(i+1,j+1)$. Because the connectivity properties that we
calculated depend only on the graph structure of the
lattice, we embedded the graph of each lattice in the square
lattice plus diagonals described above. Depending on the
lattice being modeled, some of the bonds in the underlying
lattice are closed with probability one, and some are open
with probability one, with the vertices identified in the
embedded lattice. The graph structures of all lattices appearing
in this paper were modeled in this way.

\section{\label{sec:series} Series Expansions \label{sec:expansions} }

\begin{figure}
  \vskip 10 pt
  \includegraphics[width=0.75\columnwidth]{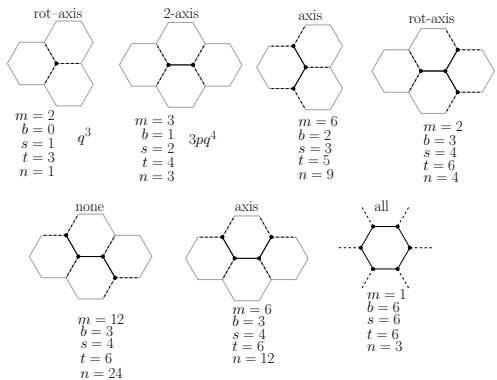}
      \caption{\label{fig:theta_hex_b} 
     Clusters on the hexagonal lattice contributing to $\theta$ to
  sixth order in $q$. Symmetries and multiplicities $m$ are:
  none--$12$, one axis--$6$, rotation--$4$, two axes--$3$,
  rotation and axis--$2$, all--$1$. The number of contributing
 clusters per figure is given by $n=ms/2$, the factor of $1/2$ accounting
 for the fact that each site only supports half the rotations. The
 contribution to $\theta$ is then $-ms(1-q)^bq^t/2$.
  }
\end{figure}

We follow the ideas of the perimeter method
\cite{domb1959,PhysRev.122.77,BleaseA} to compute high-density
 series expansions of $\theta(p)$ and $\pi(p)$ listed in
\tabr{tab:series}.  Here we
discuss the method for computing the series for $\theta(p)$,
but our method for computing $\pi(p)$ is similar.  Here, a cluster is
any connected subgraph that contains at least one vertex. We
denote by $\crt$ the collection of all finite clusters that
include the vertex at the origin, and by $\cfr$ the
partition of $\crt$ induced by equivalence under
translation, in other words the collection of free
clusters. Choosing an enumeration $\alpha_j$ of clusters in
$\cfr$, the probability that a randomly selected site is in
the infinite open cluster is easily seen to be
\begin{equation*}
 \theta(p) = 1 - \sum_{j=1}^\infty s_j (1-q)^{b_j} q^{t_j},
\end{equation*}
where $s_j$ is the number of sites and $b_j$ the number of
bonds in cluster $\alpha_j$, and $t_j$ is the number of perimeter
bonds (bonds adjacent to cluster $\alpha_j$). Clearly, we can find
the series expansion in $q$ by enumerating the clusters in
an order that is non-decreasing in $t_j$. For instance, on
the hexagonal lattice, only the cluster consisting of an
isolated site has a perimeter $t$ of size less than or equal to
$3$ (see \figr{fig:theta_hex_b}), so that to lowest
non-trivial order $\theta(p)=1-q^3$. Percolation theory is a
difficult subject precisely because the full enumeration is
difficult. The series for $\theta_\triangle$ (for bond
percolation) was calculated by machine to high order in
Ref.~\onlinecite{BleaseA}. Although tables of cluster numbers have
been published, all the others that we are aware of are either for
site percolation or for bond percolation with cluster size
measured by the number of bonds rather than sites (neither of which
can be mapped to our problem.)

For our results we counted a few small clusters by hand,
which is not so difficult. In practice however, we find that
it is also essential to categorize the clusters by
symmetry. As an example, the clusters contributing to
$\theta$ on the hexagonal lattice to sixth order in $q$ are
shown in \figr{fig:theta_hex_b}.

\begin{thebibliography}{27}
\expandafter\ifx\csname natexlab\endcsname\relax\def\natexlab#1{#1}\fi
\expandafter\ifx\csname bibnamefont\endcsname\relax
  \def\bibnamefont#1{#1}\fi
\expandafter\ifx\csname bibfnamefont\endcsname\relax
  \def\bibfnamefont#1{#1}\fi
\expandafter\ifx\csname citenamefont\endcsname\relax
  \def\citenamefont#1{#1}\fi
\expandafter\ifx\csname url\endcsname\relax
  \def\url#1{\texttt{#1}}\fi
\expandafter\ifx\csname urlprefix\endcsname\relax\def\urlprefix{URL }\fi
\providecommand{\bibinfo}[2]{#2}
\providecommand{\eprint}[2][]{\url{#2}}

\bibitem[{\citenamefont{Horodecki et~al.}()\citenamefont{Horodecki, Horodecki,
  Horodecki, and Horodecki}}]{HHH08}
\bibinfo{author}{\bibfnamefont{R.}~\bibnamefont{Horodecki}},
  \bibinfo{author}{\bibfnamefont{P.}~\bibnamefont{Horodecki}},
  \bibinfo{author}{\bibfnamefont{M.}~\bibnamefont{Horodecki}},
  \bibnamefont{and}
  \bibinfo{author}{\bibfnamefont{K.}~\bibnamefont{Horodecki}},
  \bibinfo{journal}{Rev. Mod. Phys.} , \bibinfo{note}{in press},
  \eprint{arXiv:quant-ph/0702225v2}.

\bibitem[{\citenamefont{Cirac et~al.}(1997)\citenamefont{Cirac, Zoller, Kimble,
  and Mabuchi}}]{CZKM97}
\bibinfo{author}{\bibfnamefont{J.~I.} \bibnamefont{Cirac}},
  \bibinfo{author}{\bibfnamefont{P.}~\bibnamefont{Zoller}},
  \bibinfo{author}{\bibfnamefont{H.~J.} \bibnamefont{Kimble}},
  \bibnamefont{and} \bibinfo{author}{\bibfnamefont{H.}~\bibnamefont{Mabuchi}},
  \bibinfo{journal}{Phys. Rev. Lett.} \textbf{\bibinfo{volume}{78}},
  \bibinfo{pages}{3221} (\bibinfo{year}{1997}),
  \eprint{arXiv:quant-ph/9611017v1}.

\bibitem[{\citenamefont{Boozer et~al.}(2007)\citenamefont{Boozer, Boca, Miller,
  Northup, and Kimble}}]{BBMNK07}
\bibinfo{author}{\bibfnamefont{A.~D.} \bibnamefont{Boozer}},
  \bibinfo{author}{\bibfnamefont{A.}~\bibnamefont{Boca}},
  \bibinfo{author}{\bibfnamefont{R.}~\bibnamefont{Miller}},
  \bibinfo{author}{\bibfnamefont{T.~E.} \bibnamefont{Northup}},
  \bibnamefont{and} \bibinfo{author}{\bibfnamefont{H.~J.}
  \bibnamefont{Kimble}}, \bibinfo{journal}{Phys. Rev. Lett.}
  \textbf{\bibinfo{volume}{98}}, \bibinfo{eid}{193601}
  (pages~\bibinfo{numpages}{4}) (\bibinfo{year}{2007}),
  \eprint{arXiv:quant-ph/0702248v1},
  \urlprefix\url{http://link.aps.org/abstract/PRL/v98/e193601}.

\bibitem[{\citenamefont{Ekert}(1991)}]{E91}
\bibinfo{author}{\bibfnamefont{A.~K.} \bibnamefont{Ekert}},
  \bibinfo{journal}{Phys. Rev. Lett.} \textbf{\bibinfo{volume}{67}},
  \bibinfo{pages}{661} (\bibinfo{year}{1991}).

\bibitem[{\citenamefont{Bennett et~al.}(1993)\citenamefont{Bennett, Brassard,
  Cr\'epeau, Jozsa, Peres, and Wootters}}]{BBC+93}
\bibinfo{author}{\bibfnamefont{C.~H.} \bibnamefont{Bennett}},
  \bibinfo{author}{\bibfnamefont{G.}~\bibnamefont{Brassard}},
  \bibinfo{author}{\bibfnamefont{C.}~\bibnamefont{Cr\'epeau}},
  \bibinfo{author}{\bibfnamefont{R.}~\bibnamefont{Jozsa}},
  \bibinfo{author}{\bibfnamefont{A.}~\bibnamefont{Peres}}, \bibnamefont{and}
  \bibinfo{author}{\bibfnamefont{W.~K.} \bibnamefont{Wootters}},
  \bibinfo{journal}{Phys. Rev. Lett.} \textbf{\bibinfo{volume}{70}},
  \bibinfo{pages}{1895} (\bibinfo{year}{1993}).

\bibitem[{\citenamefont{Poppe et~al.}(2008)\citenamefont{Poppe, Peev, Maurhart,
  and (on behalf of~the Integrated European Project~SECOQC)}}]{seqoqc}
\bibinfo{author}{\bibfnamefont{A.}~\bibnamefont{Poppe}},
  \bibinfo{author}{\bibfnamefont{M.}~\bibnamefont{Peev}},
  \bibinfo{author}{\bibfnamefont{O.}~\bibnamefont{Maurhart}}, \bibnamefont{and}
  \bibinfo{author}{\bibnamefont{(on behalf of~the Integrated European
  Project~SECOQC)}}, \bibinfo{journal}{Int. J. Quant. Inf.}
  \textbf{\bibinfo{volume}{6}}, \bibinfo{pages}{209} (\bibinfo{year}{2008}),
  \eprint{arXiv:quant-ph/0701168}.

\bibitem[{\citenamefont{Cirac et~al.}(1999)\citenamefont{Cirac, Ekert, Huelga,
  and Macchiavello}}]{CEHM99}
\bibinfo{author}{\bibfnamefont{J.~I.} \bibnamefont{Cirac}},
  \bibinfo{author}{\bibfnamefont{A.~K.} \bibnamefont{Ekert}},
  \bibinfo{author}{\bibfnamefont{S.~F.} \bibnamefont{Huelga}},
  \bibnamefont{and}
  \bibinfo{author}{\bibfnamefont{C.}~\bibnamefont{Macchiavello}},
  \bibinfo{journal}{Phys. Rev. A} \textbf{\bibinfo{volume}{59}},
  \bibinfo{pages}{4249} (\bibinfo{year}{1999}),
  \eprint{arXiv:quant-ph/9803017v2}.

\bibitem[{\citenamefont{Nielsen and Chuang}(2000)}]{Nielsen}
\bibinfo{author}{\bibfnamefont{M.}~\bibnamefont{Nielsen}} \bibnamefont{and}
  \bibinfo{author}{\bibfnamefont{I.}~\bibnamefont{Chuang}},
  \emph{\bibinfo{title}{Quantum Computation and Quantum Information}}
  (\bibinfo{publisher}{Cambridge University Press},
  \bibinfo{address}{Cambridge}, \bibinfo{year}{2000}).

\bibitem[{\citenamefont{Briegel et~al.}(1998)\citenamefont{Briegel, D\"ur,
  Cirac, and Zoller}}]{BDCZ98}
\bibinfo{author}{\bibfnamefont{H.-J.} \bibnamefont{Briegel}},
  \bibinfo{author}{\bibfnamefont{W.}~\bibnamefont{D\"ur}},
  \bibinfo{author}{\bibfnamefont{J.~I.} \bibnamefont{Cirac}}, \bibnamefont{and}
  \bibinfo{author}{\bibfnamefont{P.}~\bibnamefont{Zoller}},
  \bibinfo{journal}{Phys. Rev. Lett.} \textbf{\bibinfo{volume}{81}},
  \bibinfo{pages}{5932} (\bibinfo{year}{1998}),
  \eprint{arXiv:quant-ph/9803056v1}.

\bibitem[{\citenamefont{D\"ur et~al.}(1999)\citenamefont{D\"ur, Briegel, Cirac,
  and Zoller}}]{DBCZ99}
\bibinfo{author}{\bibfnamefont{W.}~\bibnamefont{D\"ur}},
  \bibinfo{author}{\bibfnamefont{H.-J.} \bibnamefont{Briegel}},
  \bibinfo{author}{\bibfnamefont{J.~I.} \bibnamefont{Cirac}}, \bibnamefont{and}
  \bibinfo{author}{\bibfnamefont{P.}~\bibnamefont{Zoller}},
  \bibinfo{journal}{Phys. Rev. A} \textbf{\bibinfo{volume}{59}},
  \bibinfo{pages}{169} (\bibinfo{year}{1999}),
  \eprint{arXiv:quant-ph/9808065v1}.

\bibitem[{\citenamefont{Childress et~al.}(2005)\citenamefont{Childress, Taylor,
  Sorensen, and Lukin}}]{CTSL05}
\bibinfo{author}{\bibfnamefont{L.}~\bibnamefont{Childress}},
  \bibinfo{author}{\bibfnamefont{J.~M.} \bibnamefont{Taylor}},
  \bibinfo{author}{\bibfnamefont{A.~S.} \bibnamefont{Sorensen}},
  \bibnamefont{and} \bibinfo{author}{\bibfnamefont{M.~D.} \bibnamefont{Lukin}},
  \bibinfo{journal}{Phys. Rev. A} \textbf{\bibinfo{volume}{72}},
  \bibinfo{eid}{052330} (pages~\bibinfo{numpages}{16}) (\bibinfo{year}{2005}),
  \eprint{arXiv:quant-ph/0502112},
  \urlprefix\url{http://link.aps.org/abstract/PRA/v72/e052330}.

\bibitem[{\citenamefont{Hartmann et~al.}(2007)\citenamefont{Hartmann, Kraus,
  Briegel, and Dur}}]{HKBD06}
\bibinfo{author}{\bibfnamefont{L.}~\bibnamefont{Hartmann}},
  \bibinfo{author}{\bibfnamefont{B.}~\bibnamefont{Kraus}},
  \bibinfo{author}{\bibfnamefont{H.-J.} \bibnamefont{Briegel}},
  \bibnamefont{and} \bibinfo{author}{\bibfnamefont{W.}~\bibnamefont{Dur}},
  \bibinfo{journal}{Phys. Rev. A} \textbf{\bibinfo{volume}{75}},
  \bibinfo{eid}{032310} (pages~\bibinfo{numpages}{17}) (\bibinfo{year}{2007}),
  \eprint{arXiv:quant-ph/0610113},
  \urlprefix\url{http://link.aps.org/abstract/PRA/v75/e032310}.

\bibitem[{\citenamefont{\ifmmode~\dot{Z}\else \.{Z}\fi{}ukowski
  et~al.}(1993)\citenamefont{\ifmmode~\dot{Z}\else \.{Z}\fi{}ukowski,
  Zeilinger, Horne, and Ekert}}]{ZZHE93}
\bibinfo{author}{\bibfnamefont{M.}~\bibnamefont{\ifmmode~\dot{Z}\else
  \.{Z}\fi{}ukowski}},
  \bibinfo{author}{\bibfnamefont{A.}~\bibnamefont{Zeilinger}},
  \bibinfo{author}{\bibfnamefont{M.~A.} \bibnamefont{Horne}}, \bibnamefont{and}
  \bibinfo{author}{\bibfnamefont{A.~K.} \bibnamefont{Ekert}},
  \bibinfo{journal}{Phys. Rev. Lett.} \textbf{\bibinfo{volume}{71}},
  \bibinfo{pages}{4287} (\bibinfo{year}{1993}).

\bibitem[{\citenamefont{Ac\'in et~al.}(2007)\citenamefont{Ac\'in, Cirac, and
  Lewenstein}}]{Natphys.3.1745}
\bibinfo{author}{\bibfnamefont{A.}~\bibnamefont{Ac\'in}},
  \bibinfo{author}{\bibfnamefont{J.~I.} \bibnamefont{Cirac}}, \bibnamefont{and}
  \bibinfo{author}{\bibfnamefont{M.}~\bibnamefont{Lewenstein}},
  \bibinfo{journal}{Nature Physics} \textbf{\bibinfo{volume}{3}},
  \bibinfo{pages}{256} (\bibinfo{year}{2007}), \eprint{arXiv:quant-ph/0612167},
  \urlprefix\url{http://dx.doi.org/10.1038/nphys549}.

\bibitem[{\citenamefont{Perseguers et~al.}(2008)\citenamefont{Perseguers,
  Cirac, Ac\'{\i}n, Lewenstein, and Wehr}}]{perseguers:022308}
\bibinfo{author}{\bibfnamefont{S.}~\bibnamefont{Perseguers}},
  \bibinfo{author}{\bibfnamefont{J.~I.} \bibnamefont{Cirac}},
  \bibinfo{author}{\bibfnamefont{A.}~\bibnamefont{Ac\'{\i}n}},
  \bibinfo{author}{\bibfnamefont{M.}~\bibnamefont{Lewenstein}},
  \bibnamefont{and} \bibinfo{author}{\bibfnamefont{J.}~\bibnamefont{Wehr}},
  \bibinfo{journal}{Phys. Rev. A} \textbf{\bibinfo{volume}{77}},
  \bibinfo{eid}{022308} (pages~\bibinfo{numpages}{14}) (\bibinfo{year}{2008}),
  \eprint{arXiv:0708.1025},
  \urlprefix\url{http://link.aps.org/abstract/PRA/v77/e022308}.

\bibitem[{\citenamefont{Kieling and Eisert}(2008)}]{kieling2007}
\bibinfo{author}{\bibfnamefont{K.}~\bibnamefont{Kieling}} \bibnamefont{and}
  \bibinfo{author}{\bibfnamefont{J.}~\bibnamefont{Eisert}}, in
  \emph{\bibinfo{booktitle}{Quantum Percolation and Breakdown}}
  (\bibinfo{publisher}{Springer}, \bibinfo{address}{Heidelberg},
  \bibinfo{year}{2008}), \bibinfo{note}{in press}, \eprint{arXiv:0712.1836}.

\bibitem[{\citenamefont{Vidal}(1999)}]{V99}
\bibinfo{author}{\bibfnamefont{G.}~\bibnamefont{Vidal}},
  \bibinfo{journal}{Phys. Rev. Lett.} \textbf{\bibinfo{volume}{83}},
  \bibinfo{pages}{1046} (\bibinfo{year}{1999}),
  \eprint{arXiv:quant-ph/9902033}.

\bibitem[{\citenamefont{Nielsen and Vidal}(2001)}]{NV01}
\bibinfo{author}{\bibfnamefont{M.~A.} \bibnamefont{Nielsen}} \bibnamefont{and}
  \bibinfo{author}{\bibfnamefont{G.}~\bibnamefont{Vidal}},
  \bibinfo{journal}{Quantum Inf. Comput.} \textbf{\bibinfo{volume}{1}},
  \bibinfo{pages}{76} (\bibinfo{year}{2001}).

\bibitem[{\citenamefont{Grimmett}(1999)}]{GrimmettA}
\bibinfo{author}{\bibfnamefont{G.}~\bibnamefont{Grimmett}},
  \emph{\bibinfo{title}{Percolation}} (\bibinfo{publisher}{Springer-Verlag},
  \bibinfo{address}{Berlin}, \bibinfo{year}{1999}), \bibinfo{edition}{2nd} ed.

\bibitem[{\citenamefont{Stauffer and Aharony}(1992)}]{StaufferB}
\bibinfo{author}{\bibfnamefont{D.}~\bibnamefont{Stauffer}} \bibnamefont{and}
  \bibinfo{author}{\bibfnamefont{A.}~\bibnamefont{Aharony}},
  \emph{\bibinfo{title}{Introduction to Percolation Theory}}
  (\bibinfo{publisher}{Taylor \& Francis}, \bibinfo{address}{London},
  \bibinfo{year}{1992}), \bibinfo{edition}{2nd} ed.

\bibitem[{\citenamefont{Ziff and Suding}(1997)}]{ziff1997}
\bibinfo{author}{\bibfnamefont{R.~M.} \bibnamefont{Ziff}} \bibnamefont{and}
  \bibinfo{author}{\bibfnamefont{P.~N.} \bibnamefont{Suding}},
  \bibinfo{journal}{J. Phys. A} \textbf{\bibinfo{volume}{30}},
  \bibinfo{pages}{5351} (\bibinfo{year}{1997}),
  \eprint{arXiv:cond-mat/9707110}.

\bibitem[{\citenamefont{Hoshen and Kopelman}(1976)}]{hoshen1976}
\bibinfo{author}{\bibfnamefont{J.}~\bibnamefont{Hoshen}} \bibnamefont{and}
  \bibinfo{author}{\bibfnamefont{R.}~\bibnamefont{Kopelman}},
  \bibinfo{journal}{Phys. Rev. B} \textbf{\bibinfo{volume}{14}},
  \bibinfo{pages}{3438} (\bibinfo{year}{1976}).

\bibitem[{\citenamefont{Nakanishi and Stanley}(1980)}]{nakanishi1980}
\bibinfo{author}{\bibfnamefont{H.}~\bibnamefont{Nakanishi}} \bibnamefont{and}
  \bibinfo{author}{\bibfnamefont{H.~E.} \bibnamefont{Stanley}},
  \bibinfo{journal}{Phys. Rev. B} \textbf{\bibinfo{volume}{22}},
  \bibinfo{pages}{2466} (\bibinfo{year}{1980}).

\bibitem[{\citenamefont{Nishimura}(2000)}]{nishimura2000}
\bibinfo{author}{\bibfnamefont{T.}~\bibnamefont{Nishimura}},
  \bibinfo{journal}{ACM Trans. on Modeling and Computer Simulation}
  \textbf{\bibinfo{volume}{10}}, \bibinfo{pages}{348} (\bibinfo{year}{2000}).

\bibitem[{\citenamefont{Domb}(1959)}]{domb1959}
\bibinfo{author}{\bibfnamefont{C.}~\bibnamefont{Domb}},
  \bibinfo{journal}{Nature} \textbf{\bibinfo{volume}{184}},
  \bibinfo{pages}{589} (\bibinfo{year}{1959}).

\bibitem[{\citenamefont{Domb and Sykes}(1961)}]{PhysRev.122.77}
\bibinfo{author}{\bibfnamefont{C.}~\bibnamefont{Domb}} \bibnamefont{and}
  \bibinfo{author}{\bibfnamefont{M.~F.} \bibnamefont{Sykes}},
  \bibinfo{journal}{Phys. Rev.} \textbf{\bibinfo{volume}{122}},
  \bibinfo{pages}{77} (\bibinfo{year}{1961}).

\bibitem[{\citenamefont{Blease et~al.}(1978)\citenamefont{Blease, Essam, and
  Place}}]{BleaseA}
\bibinfo{author}{\bibfnamefont{J.}~\bibnamefont{Blease}},
  \bibinfo{author}{\bibfnamefont{J.~W.} \bibnamefont{Essam}}, \bibnamefont{and}
  \bibinfo{author}{\bibfnamefont{C.~M.} \bibnamefont{Place}},
  \bibinfo{journal}{J. Phys. C.} \textbf{\bibinfo{volume}{11}},
  \bibinfo{pages}{4009} (\bibinfo{year}{1978}).

\end{thebibliography}

\end{document}